\documentclass[12pt, a4paper]{article}
\usepackage{amsfonts}
\usepackage{amsmath}
\usepackage{amssymb}
\usepackage[toc]{appendix}
\usepackage[hiresbb]{graphicx}
\usepackage{caption}
\usepackage{subcaption}
\usepackage{booktabs}
\usepackage[utf8]{inputenc}
\usepackage{listings}
\usepackage{here}
\usepackage{float}
\usepackage{ascmac}
\usepackage{tikz}
\usepackage{url}
\usepackage{lipsum}
\usepackage{authblk}
\usepackage{eqnarray}
\usepackage{siunitx}
\usepackage[sectionbib,round]{natbib}

\newcommand{\be}{\begin{equation}}
\newcommand{\ee}{\end{equation}}
\newcommand{\beq}{\begin{eqnarray*}}
\newcommand{\eeq}{\end{eqnarray*}}
\def\sym#1{\ifmmode^{#1}\else\(^{#1}\)\fi}

\title{\large{\bf{Impact Evaluation on the European Privacy Laws governing generative-AI models 
--- Evidence in Relation between Internet Censorship and the Ban of ChatGPT in Italy}}}
\author{\large{\bf{Tatsuru Kikuchi}}}
\affil{\small{\it{Faculty of Economics, The University of Tokyo,}}\\
{\it{7-3-1 Hongo, Bunkyo-ku, Tokyo 113-0033 Japan}}}
\date{(\today)}
\begin{document}
\maketitle
\begin{abstract}
We proceed an impact evaluation on the European Privacy Laws governing generative-AI models, especially, focusing on the effects of the Ban of ChatGPT in Italy. We investigate on the causal relationship between Internet Censorship Data and the Ban of ChatGPT in Italy during the period from March 27, 2023 to April 11, 2023. We analyze the relation based on the hidden Markov model with Poisson emissions. We find out that the HTTP Invalid Requests, which decreased during those period, can be explained with seven-state model. Our findings shows the apparent inability for the users in the internet accesses as a result of EU regulations on the generative-AI.   
\end{abstract}
\newpage
%%%
\section{Introduction}\label{sec1}
The use of artificial intelligence in the EU will be regulated by the AI Act, the world’s first comprehensive AI law. In April 2021, the European Commission proposed the first EU regulatory framework for AI. It says that AI systems that can be used in different applications are analyzed and classified according to the risk they pose to users. The different risk levels will mean more or less regulation. The regulation, agreed in negotiations with member states in December 2023, was endorsed by MEPs with 523 votes in favor, 46 against and 49 abstentions. 

According to the European Commission, It aims to protect fundamental rights, democracy, the rule of law and environmental sustainability from high-risk AI, while boosting innovation and establishing Europe as a leader in the field. The regulation establishes obligations for AI based on its potential risks and level of impact. General-purpose AI (GPAI) systems, and the GPAI models they are based on, must meet certain transparency requirements, including compliance with EU copyright law and publishing detailed summaries of the content used for training. The more powerful GPAI models that could pose systemic risks will face additional requirements, including performing model evaluations, assessing and mitigating systemic risks, and reporting on incidents.

The European Parliament today passed its landmark AI Act – a sweeping piece of legislation targeting the risks posed by the fast-moving technology. It threatens an outright ban on artificial intelligence (AI) applications which carry unacceptable risks for the safety, livelihoods and rights of EU citizen (this includes for example cognitive behavioral manipulation, social scoring or biometric identification).

It also places significant obligations on the use of AI in ‘high risk’ applications, such as health, critical infrastructure, border control, education, justice and the everyday services relied on by European citizens. The law will apply to businesses operating in the EU and, critically, the tech giants behind the AI products used by Europeans every day.

Specifically, the ChatGPT was banned in Italy on March 27, 2023, which became the first known instance of the chatbot being blocked by a government order. Italy’s data protection authority said OpenAI, the California company that makes ChatGPT, unlawfully collected personal data from users and did not have an age-verification system in place to prevent minors from being exposed to illicit material. Because of this decision, Italy became the first government to ban ChatGPT as a result of privacy concerns. In China, North Korea, Russia and Iran, the service is unavailable because OpenAI decided not to make it accessible.

On April 11, 2023, ChatGPT maker OpenAI has restored access to its service in Italy, saying it has implemented changes to satisfy Italian regulators. OpenAI said it had “addressed or clarified” the issues raised by the Italian Data Protection Authority (or GPDP) in late March. The GPDP accused ChatGPT of unlawfully collecting users’ data and failing to prevent underage users from accessing inappropriate material, leading OpenAI to block ChatGPT in the country.

\section{Internet Censorship Data}\label{sec2}
The Internet Censorship Data was provided by the Open Observatory of Network Interference (OONI). The OONI's Measurement Aggregation Toolkit (MAT) is a tool that enables you to generate your own custom charts based on aggregate views of real-time OONI data collected from around the world. The OONI data consists of network measurements collected by OONI Probe users around the world. These measurements contain information about various types of internet censorship, such as the blocking of websites and apps around the world.

Among the Internet Censorship Data provided by OONI, we use to analyze the data which shows the HTTP Invalid Requests in Italy. Those data was taken during the period of time from October 1, 2022 to September 29, 2023. The historical data on the HTTP Invalid Requests shows that it apparently decreases the requests during the period between March 27, 2023 and April 11, 2023, which corresponds to the ban period, as shown in Figure 1. 

\begin{figure}[H]
\caption{HTTP Invalid Requests in Italy (daily count)}
\centering
\includegraphics[width=1.0\textwidth]{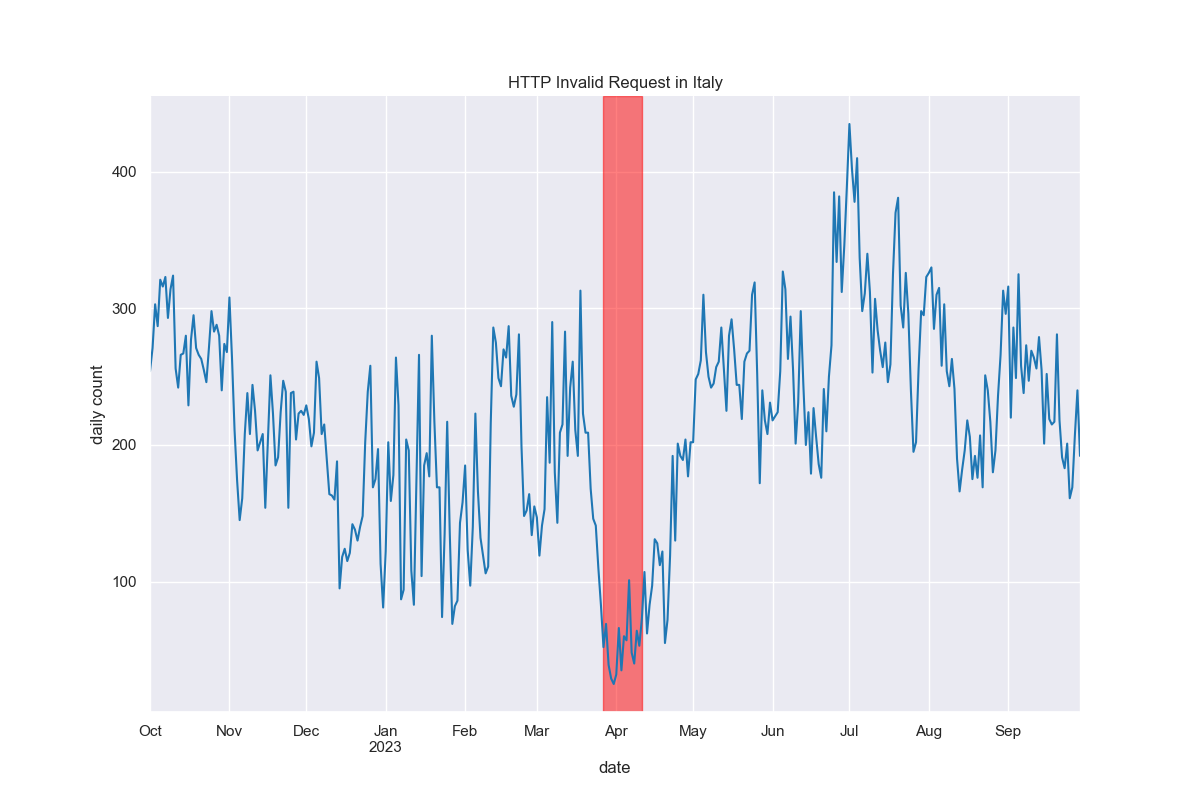}
\end{figure}

\section{Bayesian Model Selection}\label{sec4}
\subsection{Known Number of States}
We'll first consider the (perhaps unrealistic) case where the number of unobserved states is known a priori. Here, we'd assume we know there are four latent states.

We model this problem as a switching (inhomogeneous) Poisson process: at each point in time, the number of events that occur is Poisson distributed, and the rate of events is determined by the unobserved system state $z_{t}$.
\be
\lambda \sim \text{Poisson}(\lambda_{z_{t}}) \;,
\ee
where the latent states are discrete: $z_{t} = \{1, 2, 3,  4 \}$, so $\lambda_{z_{t}} = [\lambda_{1}, \lambda_{2}, \lambda_{3},  \lambda_{4} ]$ is a simple vector containing a Poisson rate for each state. To model the evolution of states over time, we'll define a simple transition model with the transition probability $p(z_{t} | z_{t-1})$. Let's say that at each step we stay in the previous state with some probability $p$, and with probability $1-p$, we transition to a different state uniformly at random. The initial state is also chosen uniformly at random, so we have:
\beq
z_{1} &\sim& \text{Categorical}\left(\left\{\frac{1}{4}, \frac{1}{4}, \frac{1}{4}, \frac{1}{4}  \right\} \right) \;,\\
z_{t} | z_{t-1} &\sim& \text{Categorical} \left( 
\begin{cases}
p & \quad \text{if} ~~ z_{t} = z_{t-1} \\
\frac{1 - p}{3} & \quad \text{otherwise}
\end{cases}
\right) \;.
\eeq
These assumptions correspond to a hidden Markov model with Poisson emissions. 

After fitting the model, we might want to reconstruct which state the model believes the system was in at each time step.

This is a posterior inference task: given the observed counts $x_{1:T}$ and model parameters (rates) $\lambda$, we want to infer the sequence of discrete latent variables, following the posterior distribution $p(z_{1:T} | x_{1:T}, \lambda)$. In a hidden Markov model, we can efficiently compute marginals and other properties of this distribution using standard message-passing algorithms. In particular, the posterior marginals method will efficiently compute the marginal probability distribution $p(Z_{t} = z_{t} | x_{1:T})$ over the discrete latent state $Z_{t}$ at each time step $t$. 

Plotting the posterior probabilities, we recover the model's "explanation" of the data as shown in Figure 2. In this (simple) case, we see that the model is usually quite confident: at most time steps it assigns essentially all probability mass to a single one of the four states. 

\begin{figure}[H]
\caption{Posterior probabilities and number of states}
\centering
\includegraphics[width=1.0\textwidth]{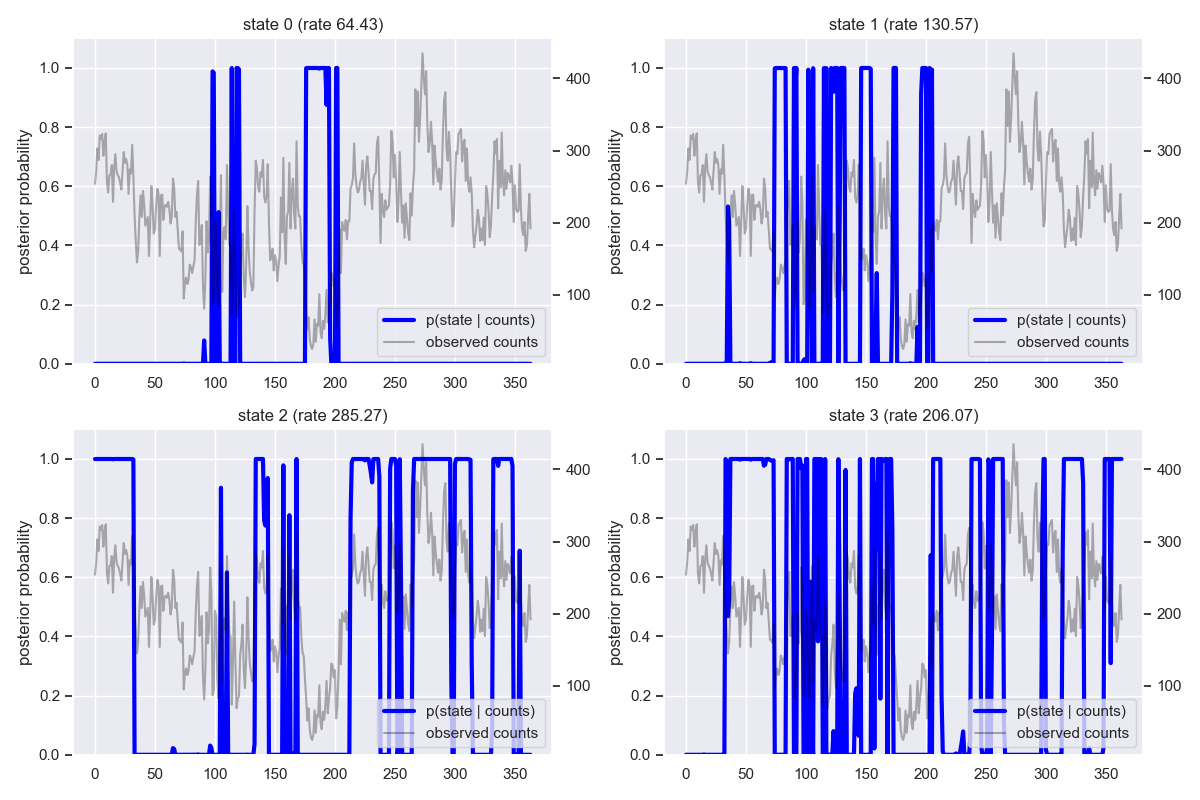}
\end{figure}

We can also visualize this posterior in terms of the rate associated with the most likely latent state at each time step, condensing the probabilistic posterior into a single explanation.

\begin{figure}[H]
\caption{The most likely latent state at each time step}
\centering
\includegraphics[width=1.0\textwidth]{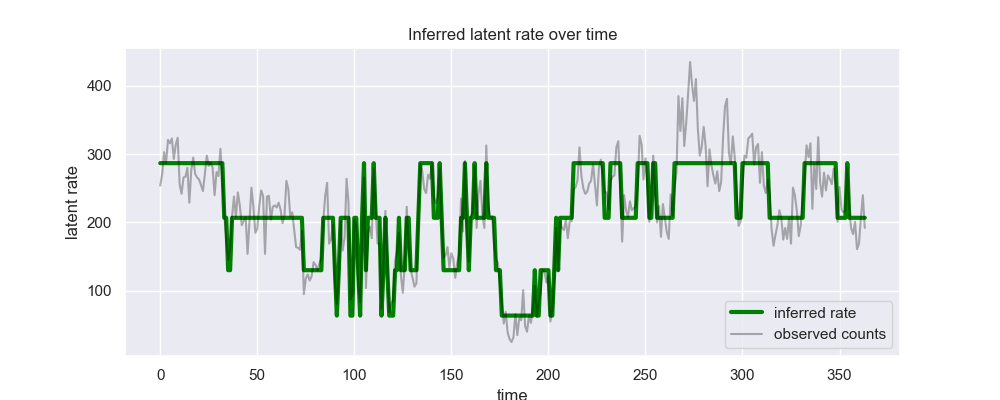}
\end{figure}

\subsection{Unknown Number of States}
In real problems, we may not know the 'true' number of states in the system we're modeling. This may not always be a concern: if you don't particularly care about the identities of the unknown states, you could just run a model with more states than you know the model will need, and learn (something like) a bunch of duplicate copies of the actual states. But let's assume you do care about inferring the 'true' number of latent states.

We can view this as a case of Bayesian model selection: we have a set of candidate models, each with a different number of latent states, and we want to choose the one that is most likely to have generated the observed data. To do this, we compute the marginal likelihood of the data under each model (we could also add a prior on the models themselves, but that won't be necessary in this analysis; the Bayesian Occam's razor turns out to be sufficient to encode a preference towards simpler models).

Unfortunately, the true marginal likelihood, which integrates over both the discrete states $z_{1:T}$ and the (vector of) rate parameters $\lambda$,
\be
p(x_{1:T}) = \int p(x_{1:T}, z_{1:T}, \lambda) dz d\lambda \;,
\ee
which is not tractable for this model. For convenience, we'll approximate it using a so-called "empirical Bayes" or "type II maximum likelihood" estimate: instead of fully integrating out the (unknown) rate parameters $\lambda$ associated with each system state, we'll optimize over their values
\be
\tilde{p}(x_{1:T}) = \max_{\lambda} \int p(x_{1:T}, z_{1:T}, \lambda) dz \;.
\ee
This approximation may overfit, {\it i.e.}, it will prefer more complex models rather than the true marginal likelihood would. We could consider more faithful approximations, e.g., optimizing a variational lower bound, or using a Monte Carlo estimator such as annealed importance sampling. 

Examining the likelihoods, we see that the (approximate) marginal likelihood tends to prefer a six-state model as shown in Figure 4. 

\begin{figure}[H]
\caption{Model Selection on latent states}
\centering
\includegraphics[width=1.0\textwidth]{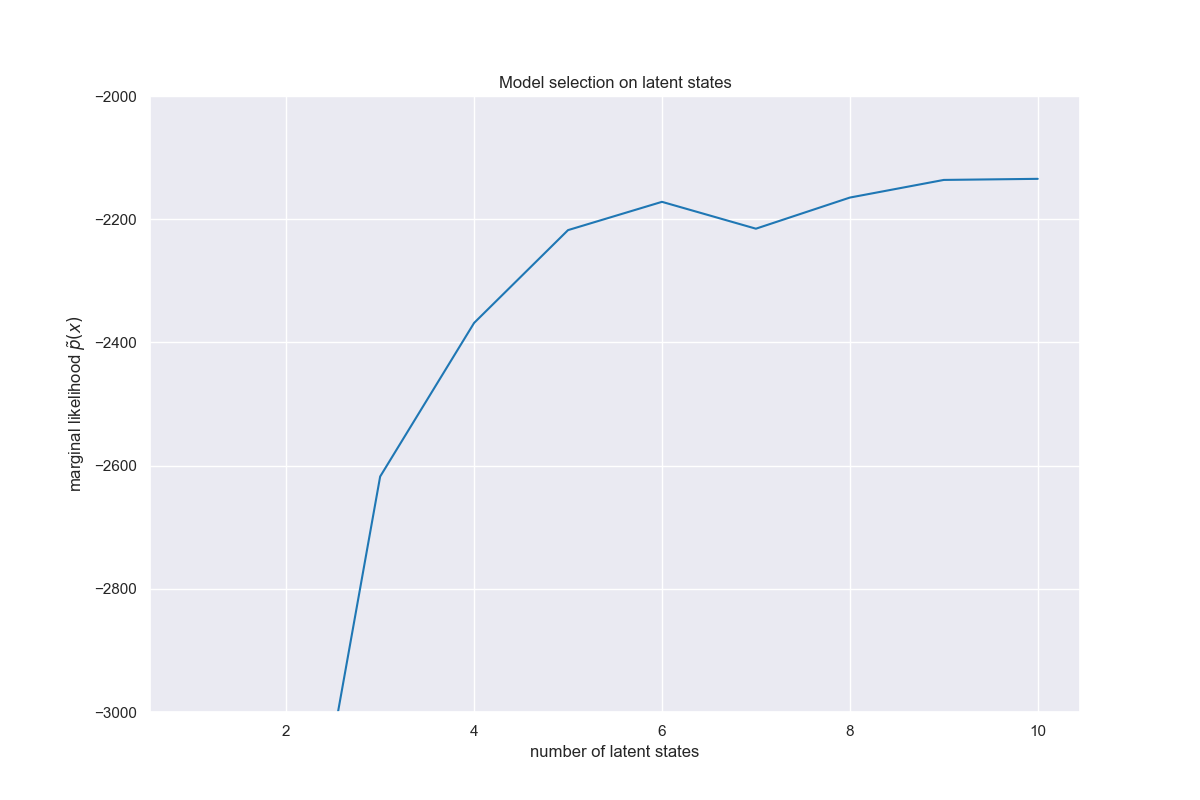}
\end{figure}

Now we proceed similarly as above. This time we'll use an extra batch dimension in trainable rates to separately fit the rates for each model under consideration. The result is shown in Figure 5.  

\begin{figure}[H]
\caption{The most likely latent state at each time step}
\centering
\includegraphics[width=1.0\textwidth]{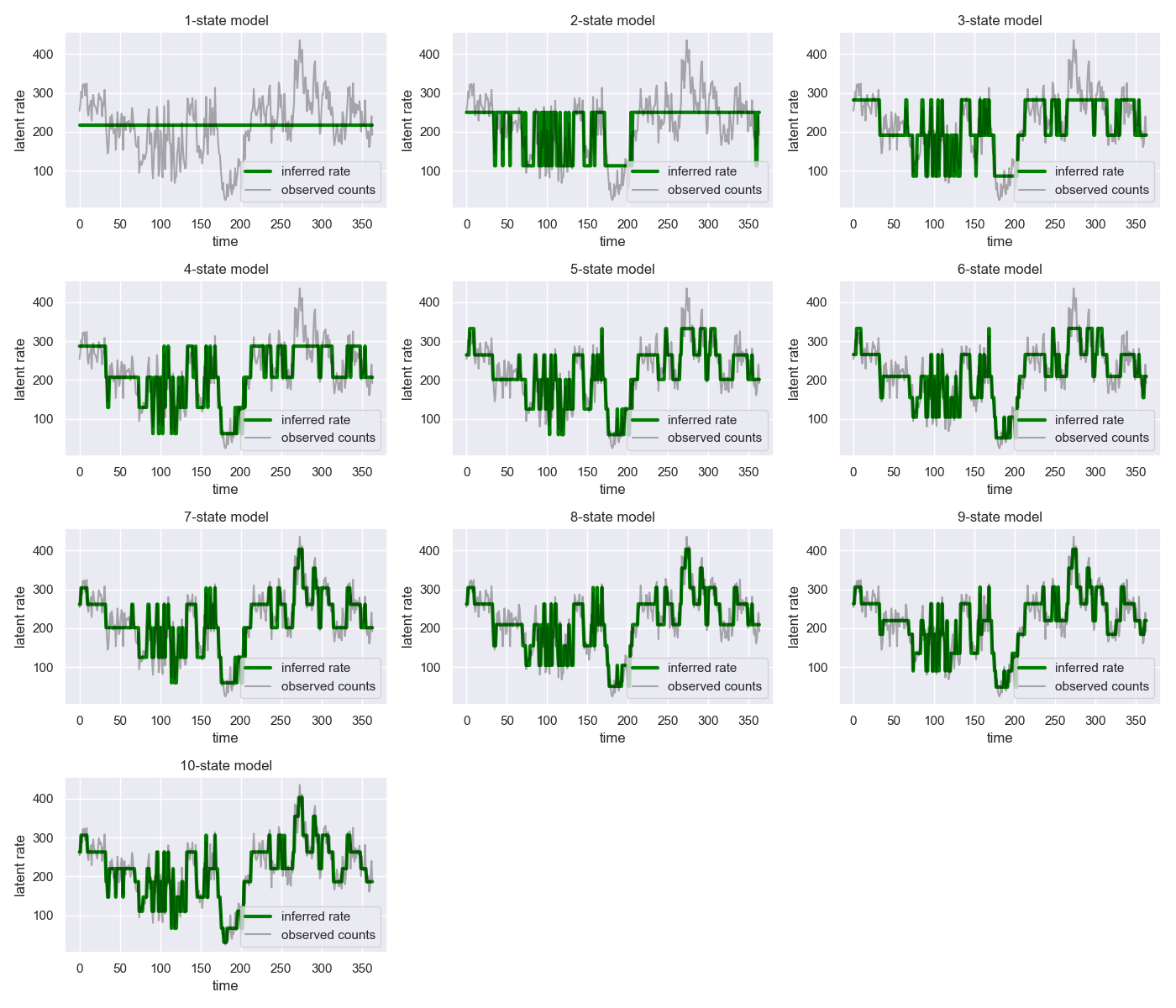}
\end{figure}

\section{Summary and Discussions}\label{sec6}
The use of artificial intelligence in the EU will be regulated by the AI Act, the world’s first comprehensive AI law. In April 2021, the European Commission proposed the first EU regulatory framework for AI. It says that AI systems that can be used in different applications are analyzed and classified according to the risk they pose to users. The different risk levels will mean more or less regulation. 

Specifically, the ChatGPT was banned in Italy on March 27, 2023, which became the first known instance of the chatbot being blocked by a government order. Italy’s data protection authority said OpenAI, the California company that makes ChatGPT, unlawfully collected personal data from users and did not have an age-verification system in place to prevent minors from being exposed to illicit material. Because of this decision, Italy became the first government to ban ChatGPT as a result of privacy concerns. 

We have proceed an impact evaluation on the European Privacy Laws governing generative-AI models, especially, focusing on the effects of the Ban of ChatGPT in Italy. Especially, we have investigate on the causal relationship between Internet Censorship Data and the Ban of ChatGPT in Italy during the period from March 27, 2023 to April 11, 2023. We have  analyzed the relation based on the hidden Markov model with Poisson emissions. 

As a result of our analysis, we finf out that the HTTP Invalid Requests, which decreased during those period, can be explained with seven-state model. Our findings shows the apparent inability for the users in the internet accesses as a result of EU regulations on the generative-AI.

\end{document}